\documentclass[aps,nofootinbib,twocolumn,prd,eqsecnum,showpacs,showkeys,preprintnumbers]{revtex4-1}
\usepackage[caption=false]{subfig}
\usepackage{graphicx}
\usepackage{amsmath}
\usepackage{amsfonts}
\usepackage{amssymb}
\usepackage{color}
\usepackage{bm}
\usepackage{mathrsfs}
\usepackage{epstopdf}
\usepackage{url}
\usepackage{footnote}
\usepackage{textcomp}

\makeatletter
\newcommand*{\rom}[1]{\expandafter\@slowromancap\romannumeral #1@}
\makeatother

\begin{document}

\title{Cosmological singularities in Born-Infeld determinantal gravity}
\author{Mariam Bouhmadi-L\'{o}pez$^{1,2,3,4}$}
\email{{\mbox{mbl@ubi.pt. On leave of absence from UPV and IKERBASQUE.}}}
\author{Che-Yu Chen$^{5,7}$}
\email{b97202056@ntu.edu.tw}
\author{Pisin Chen$^{5,6,7,8}$}
\email{pisinchen@phys.ntu.edu.tw}
\date{\today}

\affiliation{
${}^1$Departamento de F\'{i}sica, Universidade
da Beira Interior, 6200 Covilh\~a, Portugal\\
${}^2$Centro de Matem\'atica e Aplica\c{c}\~oes
da Universidade da Beira Interior (CMA-UBI)\\
${}^3$Department of Theoretical Physics, University of the Basque Country
UPV/EHU, P.O. Box 644, 48080 Bilbao, Spain\\
${}^4$IKERBASQUE, Basque Foundation for Science, 48011, Bilbao, Spain\\
${}^5$Department of Physics, National Taiwan University, Taipei, Taiwan 10617\\
${}^6$LeCosPA, National Taiwan University, Taipei, Taiwan 10617\\
${}^7$Graduate Institute of Astrophysics, National Taiwan University, Taipei, Taiwan 10617\\
${}^8$Kavli Institute for Particle Astrophysics and Cosmology, SLAC National Accelerator Laboratory, Stanford University, Stanford, CA 94305, U.S.A.
}

\begin{abstract}
The Born-Infeld determinantal gravity has been recently proposed as a way to smooth the Big Bang singularity. This theory is formulated on the Weitzenb\"{o}ck space-time and the teleparallel representation is used instead of the standard Riemannian representation. We find that although this theory is shown to be singularity-free for certain region of the parameter space in which the divergence of the Hubble rate at the high energy regime is substituted by a de-Sitter stage or a bounce in a Friedmann-Lema\^itre-Robertson-Walker universe, cosmological singularities such as Big Rip, Big Bang, Big Freeze, and Sudden singularities can emerge in other regions of the configuration space of the theory. We also show that all these singular events exist even though the Universe is filled with a perfect fluid with a constant equation of state.
\end{abstract}


\maketitle

\section{Introduction}
Undeniably, the Einstein theory of general relativity (GR) has been an extremely successful theory {\color{black}for about a century} \cite{gravitation}. However, the theory is expected to break down at some point at very high energies where quantum effects are expected to become important, for example in the past expansion of the Universe where GR predicts a big bang singularity \cite{largescale}. This is one of the motivations for looking for possible modified theories of gravity which are hoped to not only be able to preserve the great achievements of GR, but also shed some light on smoothing the {\color{black}singularities} predicted in GR. {\color{black}Such} theories could be seen as effective/phenomenological approaches of a more fundamental quantum theory of gravity.

Among the plethora of approaches to extend GR, theories with a Born-Infeld inspired action for the gravitational fields are {\color{black}attractive in that they offer the possibility of smoothing certain singularities in GR} (see Ref.~\cite{Comelli:2004qr} and the references therein). One {\color{black}is} also {\color{black}reminded} the huge success of the original Born-Infeld theory in solving the divergence of {\color{black}the} self-energy of point-like charge in the non-linear classical electrodynamics \cite{Born:1934gh}.

Recently, one of these theories called Eddington inspired Born-Infeld (EiBI) gravity was introduced in Refs.~\cite{Vollick:2003qp,Vollick:2005gc,Banados:2010ix} with the aim of smoothing the singular states in GR. This theory was also carefully studied from cosmological and astrophysical points of view \cite{Bouhmadi-Lopez:2013lha,Scargill:2012kg,Avelino:2012ue,EscamillaRivera:2012vz,Yang:2013hsa,Du:2014jka,Wei:2014dka,Delsate:2012ky,Pani:2011mg,Pani:2012qb,Casanellas:2011kf,Avelino:2012ge,Avelino:2012qe,Harko:2013wka,Harko:2013aya,Sham:2013cya,Makarenko:2014lxa,Makarenko:2014nca,Odintsov:2014yaa,Pani:2012qd,Bouhmadi-Lopez:2014jfa}. Like Eddington theory \cite{Eddington}, EiBI theory is equivalent to GR in vacuum but different in the presence of matter. Basically, (i) the Big Bang singularity and (ii) the singular state after the collapse of non-interacting particles are cured in the EiBI setup within the Newtonian limit \cite{Pani:2012qb}. Recently, we showed that dark energy related singularities can also be partially cured/smoothed in the EiBI theory both with respect to the physical metric (coupled to matter), and the auxiliary metric compatible with the physical connection \cite{Bouhmadi-Lopez:2014jfa} even if some singular states maintain their singular behaviours \cite{Bouhmadi-Lopez:2013lha,Bouhmadi-Lopez:2014jfa}.

The Born-Infeld determinantal gravity has been recently proposed as a way to smooth the Big Bang singularity \cite{Fiorini:2013kba} (see also Ref.~\cite{Ferraro:2009zk}). This theory is constructed within the Weitzenb\"{o}ck space-time and it ensures second order equations of motion of the vielbein field. In Ref.~\cite{Fiorini:2013kba}, regular cosmological solutions were obtained for some regions of the parameter space. In fact, the possible divergence of the Hubble rate at high energies is substituted by a de-Sitter phase or a bounce in a Friedmann-Lema\^itre-Robertson-Walker (FLRW) universe. In this paper we will show that although this theory is singular-free in some regions of the parameter space, cosmological singularities such as the Big Rip, Big Freeze, and Sudden singularities can {\color{black}still} emerge in other regions of the parameter space. This applies even to the Big Bang singularity. We show as well that these singular behaviours appear in a homogeneous and isotropic universe filled with a perfect fluid whose equation of state is constant.

In this paper, we will characterise the cosmological singularities by the behaviour of the Hubble rate and its cosmic time derivative at the singular event \cite{Nojiri:2005sx}:

\begin{itemize}

\item Big Rip singularity{\color{black}: It} happens at a finite cosmic time with an infinite scale factor where the Hubble parameter and its cosmic time derivative diverge \cite{Starobinsky:1999yw,Caldwell:2003vq,Caldwell:1999ew,Carroll:2003st,Chimento:2003qy,Dabrowski:2003jm,GonzalezDiaz:2003rf,GonzalezDiaz:2004vq}.

\item Sudden singularity{\color{black}: It} takes place at a finite cosmic time with a finite scale factor, where the Hubble parameter remains finite but its cosmic time derivative diverges \cite{Barrow:2004xh,Gorini:2003wa,Nojiri:2005sx}.

\item Big Freeze singularity{\color{black}: It} happens at a finite cosmic time with a finite scale factor where the Hubble parameter and its cosmic time derivative diverge \cite{BouhmadiLopez:2007qb,BouhmadiLopez:2006fu,Nojiri:2005sx,Nojiri:2004pf,Nojiri:2005sr}.

\item Type \rom{4} singularity{\color{black}: It} occurs at a finite cosmic time with a finite scale factor where the Hubble parameter and its cosmic time derivative remain finite, but higher cosmic time derivatives of the Hubble parameter still diverge \cite{BouhmadiLopez:2006fu,Nojiri:2005sx,Nojiri:2004pf,Nojiri:2005sr,Nojiri:2008fk,Bamba:2008ut}.

\end{itemize}

{\color{black}This} paper is outlined as follows. In section~II, we briefly  review the background of the Born-Infeld determinantal gravity \cite{Fiorini:2013kba}, including its action, the background of the teleparallel representation and the low energy limit of the theory. In section~III, we focus on the cosmological solutions of the theory by analysing the evolution of the Hubble rate and its cosmic time derivative as functions of the energy density of a perfect fluid with a constant equation of state. We show that within some regions of the parameter space, divergences of the Hubble rate and its cosmic time derivative can occur at a finite cosmic time, which imply the existence of the cosmological singularities mentioned above. We finally present our conclusions in section~IV.

\section{A Brief Introduction of the theory}\label{sectII}

In Ref.~\cite{Fiorini:2013kba}, a Born-Infeld determinantal gravity within a teleparallel representation was constructed aiming to get regular cosmological solutions within a FLRW space-time. The gravitational action reads \cite{Fiorini:2013kba}
\begin{equation}
I=\frac{\lambda}{2}\int d^Dx\Big[\sqrt{|g_{\mu\nu}+2\lambda^{-1}F_{
\mu\nu}|}-\sqrt{|g_{\mu\nu}|}\Big],
\label{action}
\end{equation}
where $D$ is the dimension of the space-time and $\lambda$ is a constant with the same dimension as that of a cosmological constant (in this paper, we will work with Planck units $8\pi G=1$ and set the speed of light to $c=1$). The most general form of $F_{\mu\nu}$ is
\begin{equation}
F_{\mu\nu}=\alpha {S_\mu}^{\lambda\rho}T_{\nu\lambda\rho}+\beta {S_{\lambda\mu}}^\rho {T^\lambda}_{\nu\rho}+\gamma g_{\mu\nu}T,
\label{F}
\end{equation}
where $\alpha$, $\beta$ and $\gamma$ are dimensionless constants. $T$ is the Weitzenb\"{o}ck invariant
\begin{equation}
T\equiv {S_\rho}^{\mu\nu}{T^\rho}_{\mu\nu},
\end{equation}
and ${S_\rho}^{\mu\nu}$ is defined by (see Refs.~\cite{Hehl:1976kj,tele})
\begin{eqnarray}
{S_\rho}^{\mu\nu}\equiv\frac{1}{4}({T_\rho}^{\mu\nu}-{T^{\mu\nu}}_\rho+{T^{\nu\mu}}_\rho)+\frac{1}{2}\delta^\nu_\rho{T_\sigma}^{\sigma\mu}-\frac{1}{2}\delta^\mu_\rho{T_\sigma}^{\sigma\nu}.\nonumber \\
\end{eqnarray}
Note that ${T_\rho}^{\mu\nu}$ are the components of the Weitzenb\"{o}ck torsion, which are defined through the Weitzenb\"{o}ck connection ${\Gamma^\rho}_{\mu\nu}={e^\rho}_a\partial_\nu {e^a}_\mu$ as ${T^\rho}_{\mu\nu}\equiv{\Gamma^\rho}_{\nu\mu}-{\Gamma^\rho}_{\mu\nu}$. In the Weitzenb\"{o}ck representation, the dynamical field is the vielbein $e^a$ rather than the metric $g_{\mu\nu}$ \cite{Hehl:1976kj,tele}. The metric relates with the vielbein through the following formula
\begin{equation}
g_{\mu\nu}=\eta_{ab}{e^a}_\mu{e^b}_\nu.
\end{equation}

Note that the relation between the standard Riemannian version of GR and the teleparallel version of GR is ensured by the equation
\begin{equation}
T=-R+2e^{-1}\partial_{\nu}(e{T_\sigma}^{\sigma\nu}),
\end{equation}
where $R$ is the Ricci scalar constructed within the standard Riemannian representation and $e$ is the determinant of the vielbein field (which is equal to $\sqrt{|g_{\mu\nu}|}$) \cite{Hehl:1976kj,tele}.

According to the action \eqref{action} and the definition of $F_{\mu\nu}$ \eqref{F}, the low energy limit of this theory ($|\lambda|\rightarrow\infty$) recovers GR as long as $Tr({F_\mu}^\nu)=T$, which is ensured by imposing the algebraic relation
\begin{equation}
\alpha+\beta+D\gamma=1.
\label{criteria}
\end{equation}
The theory \eqref{action} will contain more solutions than the one provided by the Einstein-Hilbert action given the freedom on the choice of the parameters $\alpha$, $\beta$ and $\gamma$. We will assume that the relation \eqref{criteria} is fulfilled from now on to guarantee the recovery of GR at the low energy limit.

\section{Born-Infeld determinantal gravity and cosmological singularities}

From now on, we will focus on a cosmological symmetry; i.e., we fix $D=4$ and choose a spatially flat FLRW space-time:
\begin{equation}
ds^2=dt^2-a^2(t)(dx^2+dy^2+dz^2),
\label{metric}
\end{equation}
where $t$ is the cosmic time and $a(t)$ is the scale factor. In addition, $e^a=diag(1,a(t),a(t),a(t))$. We assume such a universe filled with a perfect fluid with energy density $\rho$ and pressure $p$. After some lenghtly calculations, {\color{black}one obtains} the modified Friedmann equation \cite{Fiorini:2013kba}
\begin{equation}
\frac{\sqrt{1-BH^2}}{\sqrt{1-AH^2}}[1+2BH^2-3ABH^4]-1=\frac{2\rho}{\lambda},
\label{field equation}
\end{equation}
where $H$ is the Hubble parameter, $A=6(\beta+2\gamma)/\lambda$ and $B=2(2\alpha+\beta+6\gamma)/\lambda$. The relation \eqref{criteria} implies
\begin{equation}
A+3B=\frac{12}{\lambda}.
\label{criteria AB}
\end{equation}
Eq.~\eqref{field equation} is the modified Friedmann equation where $H\equiv\dot a/a$ and the overdot denotes the cosmic time derivative. 
Notice that the energy density is conserved in this type of theories; i.e., $\dot{\rho}+3H(p+\rho)=0$. The previous Friedmann equation assumes an arbitrary equation of state for the perfect fluid. At low energies ($\rho\rightarrow 0$ or  $\rho\ll \lambda$), this theory recovers GR {\color{black}(where Eq.~\eqref{criteria AB} is invoked)}:
\begin{eqnarray}
H^2&=&\frac{4}{\lambda(A+3B)}\rho-\frac{3\lambda(A^2-6AB-3B^2)}{432}\rho^2\nonumber\\
&&+O^3(\rho)\nonumber\\
&=&\frac{\rho}{3}-\frac{3\lambda(A^2-6AB-3B^2)}{432}\rho^2+O^3(\rho).
\label{low energy limit}
\end{eqnarray}

The Raychaudhuri equation reads
\begin{equation}
\dot{H}=-\frac{3}{2}(\rho+p)\frac{dH^2}{d\rho},
\label{Hdot}
\end{equation}
where 
\begin{equation}
\frac{dH^2}{d\rho}=\frac{4\sqrt{(1-AH^2)^3(1-BH^2)}}{\lambda K(H^2)},
\label{field equation2}
\end{equation}
and
\begin{eqnarray}
K(H^2)&=&A+3B-14ABH^2-6B^2H^2\nonumber\\
&&+9A^2BH^4-12A^2B^2H^6\nonumber\\
&=&\frac{12}{\lambda}-14ABH^2-6B^2H^2\nonumber\\
&&+9A^2BH^4-12A^2B^2H^6.
\label{K}
\end{eqnarray}
 
This theory not only recovers GR in the low energy limit, but also provides some interesting modifications at high energies. In Ref.~\cite{Fiorini:2013kba}, the author gave some examples where the Big Bang singularity in GR may be cured by a de-Sitter phase or a bounce for some values of $A$ and $B$. In this note we will provide a general analysis on the behaviour of $H^2$ and $\dot{H}$ for different regions of the parameter space. We will show that even if this theory is {\color{black}singularity-free} under some specific prior assumptions of the parameters, cosmological singularities such as the Big Rip, Big Freeze, and Sudden singularity {\color{black}still occur} in some regions of the parameter space. This applies even to the Big Bang singularity {\color{black}itself}. Besides, we will also show that these singularities exist even if the Universe is filled with a perfect fluid with the simplest equation of state $p=w\rho$, where $w$ is a constant.

According to the derivation of the field equation \eqref{field equation}{\color{black},} we find that there are two restrictions on the parameters of the theory: 
\begin{eqnarray}
1-AH^2>0\rightarrow\frac{1}{H^2}>A,\nonumber\\
1-BH^2>0\rightarrow\frac{1}{H^2}>B.
\label{restriction}
\end{eqnarray}
The reason is the following: the components of the tensor $g_{\mu\nu}+2\lambda^{-1}F_{\mu\nu}$, given {\color{black}in} the action~\eqref{action}, are
\begin{eqnarray}
g_{00}+2\lambda^{-1}F_{00}&=&1-AH^2,\nonumber\\
g_{ij}+2\lambda^{-1}F_{ij}&=&-a^2(t)(1-BH^2)\delta_{ij}.
\end{eqnarray}
Therefore, the restrictions shown in \eqref{restriction} ensure the consistency of the Lorentzian signature between this tensor and the physical metric $g_{\mu\nu}$, that is
%, $(+,-,-,-)$ shown in Eq.~\eqref{metric} in such a way that 
the signature of $g_{\mu\nu}+2\lambda^{-1}F_{\mu\nu}$ does not change when recovering GR (i.e. $|\lambda|\rightarrow \infty$).

According to the conditions \eqref{restriction}, one can see that the Hubble rate may {\color{black}become large} if both $A$ and $B$ are negative, otherwise, the Hubble rate has an upper bound and may imply a regular cosmological behaviour. Thus, whether the parameter $\lambda$ is positive or not plays a {\color{black}crucial} role. For the sake of convenience, we will split the discussion about the issue of singularities in this theory into two cases: $\lambda>0$ and $\lambda<0$.

\subsection{$\lambda>0$}
\begin{figure}[!h]
\graphicspath{{fig/}}
\includegraphics[scale=0.8]{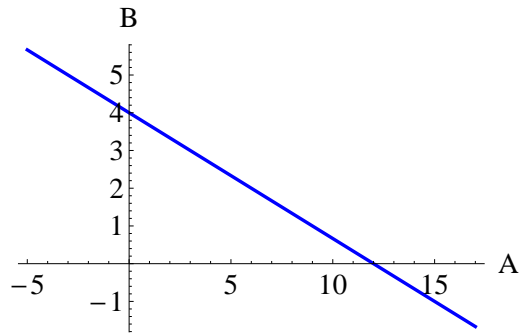}
\caption{The allowed parameter space for $A$ and $B$ is shown by the solid blue line for positive $\lambda$. The unit of this figure is $1/\lambda$.}
\label{AB positive}
\end{figure}

We start {\color{black}with} assuming $\lambda$ positive. This assumption was made in Ref.~\cite{Fiorini:2013kba} to exhibit the ability of the Born-Infeld determinantal theory to avoid singular GR behaviour, namely the Big Bang. In FIG.~\ref{AB positive}, we show the allowable parameter space of $A$ and $B$ according to the criteria \eqref{criteria AB} for a positive $\lambda$ as a blue line. One can see that the solid blue line does not cross the fourth quadrant. This means that if $\lambda>0$, at least one of the following inequalities is satisfied:
\begin{equation}
H^2<\frac{1}{A}\textrm{ or }H^2<\frac{1}{B}.
\end{equation}
Therefore, the singularities in which the Hubble rate blows up such as the Big Bang, Big Rip, Little Rip, Big Freeze can all be cured.

However, even if the Hubble rate is restricted to be finite, the cosmic time derivative of the Hubble rate may diverge at a finite cosmic time and at a finite scale factor, thus a sudden singularity may happen for $\lambda>0$. According to Eqs.~\eqref{Hdot}, \eqref{field equation2} and \eqref{K}, the cosmic time derivative blows up as long as a positive solution for $K(H^2)=0$ exists. If this divergence happens when the Hubble rate, the scale factor, and the cosmic time are finite, it implies a sudden singularity. We next show this is indeed the case. For example, if we assume $A=0$ and therefore $B=4/\lambda$, the modified field equation becomes:
\begin{equation}
\sqrt{1-\frac{4}{\lambda}H^2}\Big[1+\frac{8}{\lambda}H^2\Big]=1+\frac{2\rho}{\lambda},
\label{field equation 3}
\end{equation}
and 
\begin{equation}
K(H^2)=12-\frac{96}{\lambda}H^2.
\end{equation}
Thus, $\dot H$ diverges when $H^2=\lambda/8$ and $\rho=\rho_{s_1}$ where $\rho_{s_1}=(\sqrt{2}-1)\lambda/2$ (cf. Eq.~\eqref{K}). Any matter content whose pressure is finite when $\rho\rightarrow\rho_{s_1}$  for a finite scale factor will imply a sudden singularity which is of a geometrical origin. The sudden singularity comes from a purely geometrical feature. A similar effect can happen on some brane-worlds models \cite{Shtanov:2002ek,Brown:2005ug,Mars:2007wy,BouhmadiLopez:2010vi,BouhmadiLopez:2009jk}. This conclusion applies to any linear equation of state like the one fulfilled by dust, radiation or even some phantom energy models. This divergence happens at a finite cosmic time because
\begin{equation}
t-t_{s_1}\propto\rho-\rho_{s_1},
\end{equation}
where $t_{s_1}$ and $\rho_{s_1}$ are the cosmic time and the energy density at the singularity. Note that the Friedmann equation \eqref{field equation 3} implies $H^2<\lambda/4$, a condition fully compatible with the value of the Hubble rate at the Sudden singularity.

\subsection{$\lambda<0$}
\begin{figure}[!h]
\graphicspath{{fig/}}
\includegraphics[scale=0.8]{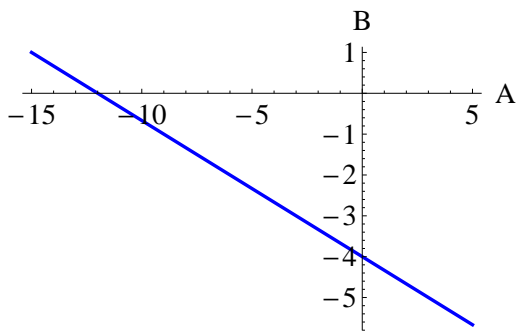}
\caption{The allowed parameter space for $A$ and $B$ is shown by the blue solid line for negative $\lambda$. The unit of this figure is $1/|\lambda|$.}
\label{AB negative}
\end{figure}

On the other hand, the behaviour of the cosmological solutions is quite different if $\lambda$ is negative. Similar to what we did in the previous subsection, in FIG.~\ref{AB negative} we show the allowable parameter space of $A$ and $B$ according to the criteria \eqref{criteria AB} for a negative $\lambda$.

The allowable parameter space of $A$ and $B$ is again shown by the solid blue line in FIG.~\ref{AB negative}. One can see that this line crosses the fourth quadrant, that is, $-12/|\lambda|\le A\le 0$. This means that $H^2$ has no upper limit in this region. Therefore, the divergence of the Hubble rate, or the occurrence of some strong dark energy singularities such as the Big Rip, Little Rip and Big Freeze could be unavoidable.

\subsubsection{$B=0$}

To exhibit the possibility of the above mentioned singular behaviours, we assume $B=0$ and $A=-12/|\lambda|$ as a first example. The modified field equation reads
\begin{equation}
H^2=\frac{|\lambda|}{12}\Big[\frac{1}{(1-\frac{2\rho}{|\lambda|})^2}-1\Big].
\end{equation}
At the low energy limit, the expansion of $H^2$ simplifies to:
\begin{equation}
H^2=\frac{\rho}{3}+\frac{\rho^2}{|\lambda|}+\frac{8\rho^3}{3\lambda^2}+O^4(\rho).
\end{equation}
Therefore, the Universe will be asymptotically Minkowski when $\rho\rightarrow 0$. However, the Hubble rate will reach infinite values as the energy density approaches its maximum value $\rho_{s_2}=|\lambda|/2$. The asymptotic behaviour of $H^2$ in this region reads
\begin{equation}
H^2\approx\frac{|\lambda|^3}{48(\rho-\rho_{s_2})^2}.
\end{equation}
Any matter content whose pressure is finite when $\rho\rightarrow\rho_{s_2}$ for a finite scale factor will indicate that this divergence happens at a finite cosmic time because
\begin{equation}
t-t_{s_2}\propto(\rho-\rho_{s_2})^2.
\end{equation}
This condition applies to any linear equation of state like the one fulfilled by dust, radiation or even some phantom energy models. Similarly, the cosmic time derivative of the Hubble rate blows up at this point.
This implies that this singularity is a Big Freeze. To the best of our knowledge this is the first time that a Big Freeze singularity \cite{BouhmadiLopez:2006fu}  can emerge from pure geometrical effects unrelated with the matter equation of state. 

\subsubsection{$A=0$}
In this subsubsection, we assume $A=0$ and $B=-4/|\lambda|$ as another guiding example to exhibit the singular behaviour of the theory. The modified field equation becomes:
\begin{equation}
\sqrt{1+\frac{4}{|\lambda|}H^2}\Big[1-\frac{8}{|\lambda|}H^2\Big]=1-\frac{2\rho}{|\lambda|},
\label{field equation n A0}
\end{equation}
and the low energy limit is
\begin{equation}
H^2=\frac{\rho}{3}-\frac{\rho^2}{3|\lambda|}+O^3(\rho).
\end{equation}
However, one can see from Eq.~\eqref{field equation n A0} that the Hubble rate diverges when $\rho\rightarrow\infty$. The asymptotic behaviour of $H^2$ near this region is
\begin{equation}
H^2\approx\frac{|\lambda|^{\frac{1}{3}}\rho^{\frac{2}{3}}}{4}.
\end{equation}
If we further assume a constant equation of state $p=w\rho$, we find that the cosmic time when the singularity happens is also finite, no matter it happens in the past or in the future:
\begin{equation}
t-t_{s_3}\propto\frac{\rho^{-\frac{1}{3}}}{1+w},
\end{equation}
where the energy density at the singularity is infinite ($t\rightarrow t_{s_3}$). Note that if $w<-1$, this singularity will happen in the future and corresponds to a Big Rip singularity; if $w>-1$, this singularity will happen in the past and corresponds to a Big Bang singularity. Notice as well that all the cosmic time derivative of $H$ blow up in both cases.

\section{conclusions}

The Born-Infeld determinantal gravity has been recently proposed as a way to smooth the Big Bang singularity \cite{Fiorini:2013kba}. This theory is constructed under the teleparallel representation and recovers GR at the low energy limit. Because of the various additional parameters in the action, this theory is characterised by its ability to construct different cosmological solutions of interest by adjusting these parameters. In Ref.~\cite{Fiorini:2013kba}, {\color{black}some regions of the parameter space} of the theory were chosen to exhibit the ability of the Born-Infeld determinantal gravity to avoid the Big Bang singularity. In this note, we generalise the analysis and find that most of the cosmological singularities such as the Big Rip, Big Freeze, and Sudden singularity {\color{black}may still} emerge on the basis of this theory{\color{black}. In fact, even the Big Bang itself may still encounter singularity in different regions of the parameter space.} More precisely, the behaviours of the solutions in which $\lambda>0$, where $\lambda$ is the main parameter of the theory, are much regular than those in which $\lambda<0$. The Hubble rate has a finite upper limit for $\lambda>0$ thus inhibit the occurrence of the Big Bang, Big Rip, Little Rip and Big Freeze singularities, while a Sudden singularity may emerge as we {\color{black}have shown} in {\color{black}S}ection.~III. On the other hand, the Hubble rate may not have a finite upper limit for $\lambda<0$, thus the divergences of the Hubble rate and the occurrence of a Big Rip, Big Bang, Big Freeze may be possible. Interestingly, we also show that these singular solutions appear even if the Universe is assumed to be only filled with a perfect fluid with a constant equation of state. We would like also to highlight that the Big Freeze singularity we found in this model is coming from pure geometrical effects and to the best of our knowledge this the first example in the literature.

\acknowledgments

The work of M.B.L. was supported by the Basque Foundation for Science IKERBASQUE and the Portuguese Agency ``Funda\c{c}\~{a}o para a Ci\^{e}ncia e Tecnologia" through an
Investigador FCT Research contract, with reference IF/01442/2013/CP1196/CT0001. She also wishes to acknowledge the hospitality of LeCosPA Center at the National Taiwan University during the completion of part of this work and the support from the Portuguese Grants PTDC/FIS/111032/2009 and PEst-OE/MAT/UI0212/2014  and the partial support from the Basque government Grant No. IT592-13 (Spain).
C.-Y.C. and P.C. are supported by Taiwan National Science Council under Project No. NSC 97-2112-M-002-026-MY3 and by Taiwan’s National Center for Theoretical Sciences (NCTS). P.C. is in addition supported by US Department of Energy under Contract No. DE-AC03-76SF00515.

\end{document}